\documentclass[lettersize,journal]{IEEEtran}

\usepackage{subfigure}
\usepackage{setspace}
\usepackage{multirow} 
\usepackage{algorithm}
\usepackage{algorithmic}
\usepackage{amsmath}
\usepackage{amssymb}
\usepackage{latexsym}
\usepackage{multirow}
\usepackage{epsfig}
\usepackage{graphics}
\usepackage{graphicx}
\usepackage{mathrsfs}
\usepackage{subfigure}
\usepackage{slashbox}
\usepackage{mathrsfs}
\usepackage{bbding}
\usepackage{cite}
\usepackage{color}
\usepackage{xcolor}
\usepackage{booktabs}
\usepackage{amssymb,mathrsfs,amsmath}
\newcommand{\bm}[1]{\mbox{\boldmath{$#1$}}}
\usepackage{cases}

\usepackage{ragged2e}

\usepackage{hyperref}

\allowdisplaybreaks [4]

\usepackage{amsthm}
\theoremstyle{plain}

\begin{document}
			
\title{Rate-Splitting Multiple Access for Intelligent Reflecting Surface-Aided Secure Transmission}
\author{Ying~Gao, Qingqing~Wu, Wen~Chen, and Derrick~Wing~Kwan~Ng,~\IEEEmembership{Fellow,~IEEE}\vspace{-10mm}
\thanks{The work of Q.~Wu was supported by SRG2020-00024-IOTSC. The work of W.~Chen was supported by National key project 2020YFB1807700 and 2018YFB1801102, by Shanghai Kewei 20JC1416502 and 22JC1404000, Pudong PKX2021-D02 and NSFC 62071296. The work of D.~W.~K.~Ng was supported by  the Australian Research Council's Discovery Project (DP210102169).  \emph{(Corresponding author: Qingqing Wu.)} }
\thanks{Y. Gao and Q. Wu are with the State Key Laboratory of Internet of Things for Smart City, University of Macau, Macau 999078, China (e-mail: yinggao@um.edu.mo; qingqingwu@um.edu.mo). W.~Chen is with the Department of Electronic Engineering, Shanghai Jiao Tong University, Shanghai 201210, China (e-mail: wenchen@sjtu.edu.cn). D.~W.~K.~Ng is with the School of Electrical Engineering and Telecommunications, University of New South Wales, NSW 2052, Australia (e-mail: w.k.ng@unsw.edu.au).}}

\maketitle

\begin{abstract}
	In this letter, we study a rate-splitting multiple access (RSMA)-based intelligent reflecting surface (IRS)-aided multi-user multiple-input single-output (MISO) secure communication system with a potential eavesdropper (Eve). Aiming to maximize the minimum secrecy rate (SR) among all the legitimate users (LUs), a design problem for jointly optimizing the transmit beamforming with artificial noise (AN), the IRS beamforming, and the secrecy common rate allocation is formulated. Since the design problem is highly non-convex with coupled optimization variables, we develop a computationally efficient algorithm based on the alternating optimization (AO) technique to solve it suboptimally. Numerical results demonstrate that the proposed design can significantly improve the max-min SR over the benchmark schemes without IRS or adopting other multiple access techniques. In particular, employing the RSMA strategy can substantially reduce the required numbers of IRS elements for achieving a target level of secrecy performance compared with the benchmark schemes.  
\end{abstract}
\vspace{-2mm}
\begin{IEEEkeywords}
	Rate-splitting multiple access, intelligent reflecting surface, physical layer security, max-min fairness. 
\end{IEEEkeywords}

\vspace{-5mm}
\section{Introduction}\label{Intro}
\IEEEPARstart{D}{ue} to the capability of reconfiguring the wireless propagation environment, intelligent reflect surface (IRS) has recently drawn considerable research attention \cite{2019_Qingqing_Joint,2020_Dong_Ergodic_Capacity,2022_Zhi_RIS_Satellite-Terrestrial}. In particular, IRS exhibits great potential in enhancing the physical layer security (PLS) for wireless networks. By intelligently adapting the IRS elements' phase shifts, the reflected and direct signals can be superimposed constructively at legitimate users (LUs) while destructively at potential eavesdroppers (Eves) \cite{2020_Qingqing_IRS_Intro}. Initial studies on IRS-aided secure wireless communications, e.g., \cite{2019_Miao_Secure,2020_Xinrong_AN}, considered a simple setting with only one LU. Notably, the simulation results in \cite{2020_Xinrong_AN} confirmed the benefit of exploiting artificial noise (AN). For a more general case of multi-antenna wireless networks with multiple LUs, two major transmission schemes, i.e., multi-user linear precoding (MU-LP) and power-domain non-orthogonal multiple access (NOMA), have been thoroughly investigated in previous works (see e.g., \cite{2020_Xianghao_secure_rank1,2021_Hehao_security,2022_Hehao_IRS_security_SWIPT,2021_Zheng_secure_NOMA}). \looseness=-1

On the other hand, rate-splitting multiple access (RSMA), bridging conventional MU-LP and NOMA, has been recently advocated as a promising transmission strategy to suppress multi-user interference \cite{2018_YiJie_RSMA}. Rate-splitting (RS) relies on the split of user messages into common and private parts, superimposed transmission from the transmitter, and successive interference cancellation (SIC) at the receivers. Particularly, the common message can not only provide extra degrees of freedom but also act as AN to enhance the quality of security provisioning \cite{2020_Hao_RSMA_robsecure}. Existing research, e.g., \cite{2018_YiJie_RSMA,2020_Hao_RSMA_robsecure,2020_Yijie_CoRSMA_maxmin} and references therein, has demonstrated that RSMA outperforms MU-LP and NOMA in several aspects such as energy efficiency (EE), robustness, user fairness, PLS, etc. 

To reap the advantages of both IRS and RSMA, some recent works have studied the amalgamation of them under different criteria, e.g., EE maximization \cite{2020_Zhaohui_IRS_RSMA_EE}, max-min fairness \cite{2021_Hao_IRS_RSMA_multiuser}, and outage performance \cite{2021_Ankur_IRS_RSMA_outage}. However, to the best of the authors' knowledge, there is no study on integrating RSMA with IRS for security performance enhancement. With the consideration of PLS, the corresponding problem formulation and resource allocation design for IRS-aided RSMA are quite different from those in e.g., \cite{2019_Qingqing_Joint,2020_Dong_Ergodic_Capacity,2022_Zhi_RIS_Satellite-Terrestrial,2020_Qingqing_IRS_Intro,2019_Miao_Secure,2020_Xinrong_AN,2020_Xianghao_secure_rank1,2021_Hehao_security,2022_Hehao_IRS_security_SWIPT,2021_Zheng_secure_NOMA}. This is not only because RSMA allows to better manage the inter-user interference by applying the principle of RS, but also because the common message can serve as AN to reduce the potential information leakage to Eves. These factors complicate the interference management in the design of IRS’s reflection, too. Therefore, the resource allocation designs proposed in e.g., \cite{2019_Qingqing_Joint,2020_Dong_Ergodic_Capacity,2022_Zhi_RIS_Satellite-Terrestrial,2020_Qingqing_IRS_Intro,2019_Miao_Secure,2020_Xinrong_AN,2020_Xianghao_secure_rank1,2021_Hehao_security,2022_Hehao_IRS_security_SWIPT,2021_Zheng_secure_NOMA} cannot be directly applied to IRS-aided RSMA. Besides, from the perspective of RSMA, it is still unknown whether the IRS can further enhance the PLS for RSMA-based systems. From the perspective of IRS, it remains unclear whether RSMA can evidently reduce the required surface size of IRS when a certain secrecy performance level is desired, which is of great practical interests for the implementation of IRSs. 

\begin{figure}[!t]
	\setlength{\abovecaptionskip}{-3pt}
	\setlength{\belowcaptionskip}{-1pt}
	\centering
	\includegraphics[width=0.31\textwidth]{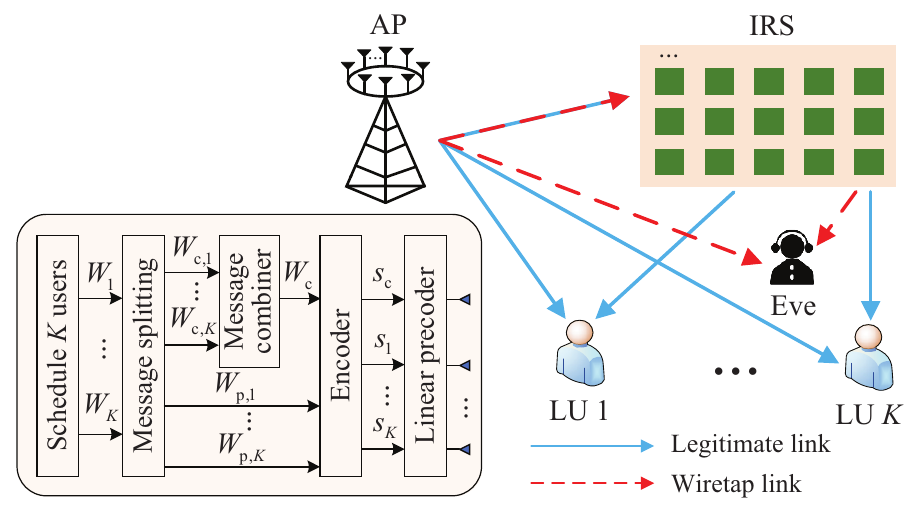}
	\caption{Illustration of IRS-aided secure transmission with $1$-layer RS.}
	\label{fig:system_model}
	\vspace{-5mm}
\end{figure}

Motivated by the above discussions, this letter investigates a RSMA-based secure downlink communication system consisting of an IRS, a multiple-antenna access point (AP), multiple single-antenna LUs, and a single-antenna Eve, as depicted in Fig. \ref{fig:system_model}. As an initial study, we consider the low-complexity $1$-layer RS strategy, where a single common stream is shared by all the LUs and a single layer of SIC is required at each LU \cite{2018_YiJie_RSMA}. A minimum secrecy rate (SR) maximization problem is formulated, where the transmit precoder and the AN covariance matrix at the AP, the IRS phase shifts, and the secrecy common rate allocation are jointly optimized. Due to the non-smoothness and non-convexity of the formulated problem, we first propose an equivalent transformation to the problem at hand to arrive a tractable reformulation. 
Then, we propose a computationally efficient alternating optimization (AO) algorithm to obtain a suboptimal solution by decomposing the resulting problem into two subproblems that are alternatingly solved until convergence is achieved. Simulation results show that remarkable security performance enhancement can be achieved by our proposed IRS-aided RSMA transmission strategy as compared to the cases without IRS or adopting MU-LP and NOMA schemes. Moreover, RSMA is more attractive to space-constrained environments as it needs fewer IRS elements than MU-LP and NOMA for achieving a given performance level. 

\emph{Notations:} 
$\left|\cdot\right|$ and $\left\|\cdot\right\|$ denote the modulus of a scalar and the Euclidean norm of a vector, respectively. $\left\|\cdot\right\|_*$ and $\left\|\cdot\right\|_2$ represent the sum of the singular values and the maximum singular value of a matrix, respectively. $[\cdot]_n$ and $[\cdot]_{i,j}$ stand for the $n$-th element of a vector and the $(i,j)$-th element of a matrix, respectively. $\mathcal {CN}\left(\mu, \sigma^2\right)$ denotes a complex Gaussian distribution with mean $\mu$ and variance $\sigma^2$. We adopt diag$(\cdot)$ to represent the diagonalization operation. The trace and conjugate transpose operators are denoted by ${\rm tr}(\cdot)$ and $(\cdot)^H$, respectively. $\bm S  \succeq \bm 0$ indicates that matrix $\bm S$ is positive semidefinite. ${\rm rank}(\cdot)$ is the rank of a matrix. 
$\nabla_{\mathbf x}f\left(\mathbf x,\mathbf y\right) $ denotes the partial gradient of function $f\left(\mathbf x,\mathbf y\right)$ with respect to (w.r.t.) vector $\mathbf x$.  \looseness=-1 

\vspace{-1mm} 
\section{System Model and Problem Formulation}
As illustrated in Fig. \ref{fig:system_model}, an IRS equipped with $N$ reflecting elements is deployed in a secure communication system, where an $M$-antenna AP serves $K$ single-antenna LUs in the presence of a single-antenna Eve\footnote{Due to the space limitation, we restrict our consideration to the case with only one Eve. However, it should be noted that the proposed algorithm can be easily extended to the case with multiple Eves.}. Let $\mathcal N\triangleq\{1,\cdots,N\}$ and $\mathcal K\triangleq \{1,\cdots,K\}$ be the sets of reflecting elements and LUs, respectively. Denoted by $\mathbf\Theta = \text{diag}\left(e^{\jmath\theta_1},\cdots,e^{\jmath\theta_N}\right)$ the reflection-coefficient matrix at the IRS, with $\theta_n\in[0,2\pi)$ representing the phase shift introduced by the $n$-th IRS element. Let $\mathbf G\in\mathbb C^{N\times M}$, $\mathbf h_{d,k}^H\in\mathbb C^{1\times M}$, $\mathbf h_{d,\rm e}^H\in\mathbb C^{1\times M}$, $\mathbf h_{r,k}^H\in\mathbb C^{1\times N}$, and  $\mathbf h_{r,\rm e}^H\in\mathbb C^{1\times N}$ denote the channel coefficients from the AP to the IRS, from the AP to LU $k$, from the AP to Eve, from the IRS to LU $k$, and from the IRS to Eve, respectively. Then, we denote the cascaded AP-IRS-LU $k$ and AP-IRS-Eve channels as $\mathbf Q_k = \text{diag}\big(\mathbf h_{r,k}^H\big)\mathbf G$ and $\mathbf Q_{\rm e} = \text{diag}\big(\mathbf h_{r,\rm e}^H\big)\mathbf G$, respectively. Suppose that the channel state information (CSI) of the direct AP-LUs/Eve channels and the cascaded IRS-related channels (instead of the individual IRS-AP/LUs/Eve channels) is perfectly known at the AP/IRS, which is sufficient for the transmit/reflect beamforming design without loss of optimality. \looseness=-1 

The 1-layer RS strategy \cite{2018_YiJie_RSMA} mentioned in Section \ref{Intro} is implemented at the AP. In addition, AN is transmitted simultaneously with the data streams to combat the potential eavesdropping by Eve. The transmitted signal is given by $\bm x = \mathbf w_{\rm c}s_{\rm c} + \sum\nolimits_{k\in\mathcal K}\mathbf w_ks_k + \mathbf z$, where $s_{\rm c} \sim\mathcal {CN} (0, 1)$ and $s_k \sim\mathcal {CN} (0, 1)$ denote the common stream for all the LUs and the private stream for LU $k$ only, respectively, which are statistically uncorrelated and precoded by the precoding vectors $\mathbf w_{\rm c} \in\mathbb C^{M\times 1}$ and $\mathbf w_k \in\mathbb C^{M\times 1}$. Besides, $\mathbf z \sim\mathcal {CN} \left(\mathbf 0, \mathbf Z \right)$ is the AN vector, with $\mathbf Z \in\mathbb H^M, \mathbf Z\succeq \mathbf 0$ being its covariance matrix. We assume that the AN is unknown to both the LUs and Eve. Then, the received signal at LU $k$ or Eve can be written as $y_j = \big(\mathbf h_{d,j}^H + \mathbf h_{r,j}^H\mathbf \Theta\mathbf G\big)\bm x + n_j  = \mathbf v^H\mathbf H_j\bm x + n_j, \ j\in\mathcal K\cup\{\rm e\}$, where $\mathbf v = \left[\mathbf u; 1\right] $ with $\mathbf u = \left[e^{\jmath\theta_1}, \cdots, e^{\jmath\theta_N} \right]^H$, $\mathbf H_j = \big[\mathbf Q_j; \mathbf h_{d,j}^H \big]$, and $n_j$ is the zero-mean additive white Gaussian noise (AWGN) with variance $\sigma_j^2$, respectively. 

Following the 1-layer RS decoding order \cite{2018_YiJie_RSMA}, the achievable rate of decoding the common stream $s_{\rm c}$ at LU $k$ or Eve in bits/second/Hertz (bps/Hz) can be expressed as $R_{\rm c,\it j} = \log_2\left( 1 + \gamma_{\rm c,\it j}\right)$, where {\small $\gamma_{\rm c,\it j}  =  \left|\mathbf v^H\mathbf H_j\mathbf w_{\rm c}\right|^2\Big/\Big(\sum_{i\in \mathcal K}\left|\mathbf v^H\mathbf H_j\mathbf w_i\right|^2 + {\rm tr}\left(\mathbf H_j^H\mathbf v\mathbf v^H\mathbf H_j\mathbf Z \right) + \sigma_j^2\Big)$}, $j\in\mathcal K\cup\{\rm e\}$. Particularly, the transmission rate of $s_{\rm c}$ shall not exceed $R_{\rm c} = \min_{k\in\mathcal K}\{R_{\rm c,\it k}\}$ so that all the LUs can successfully decode $s_{\rm c}$. After decoding $s_{\rm c}$, LU $k$ removes it from the received signal via SIC and then decodes the private stream $s_k$. Hence, the achievable rate of decoding $s_k$ at LU $k$ in bps/Hz is given by $R_{\rm p,\it k} = \log_2\left( 1 + \gamma_{\rm p,\it k}\right)$, with {\small $\gamma_{\rm p,\it k} = \left|\mathbf v^H\mathbf H_k\mathbf w_k\right|^2 \Big/\Big(\sum_{i\in \mathcal K\backslash\{k\}}\left|\mathbf v^H\mathbf H_k\mathbf w_i\right|^2 + {\rm tr}\left(\mathbf H_k^H\mathbf v\mathbf v^H\mathbf H_k\mathbf Z \right) + \sigma_k^2\Big)$}. On the other hand, to enable the common message to act as AN for degrading the SINR of each private message at Eve, we need to prevent $s_{\rm c}$ from being decoded by Eve. For this purpose, the condition $R_{\rm c,e} < R_{\rm c}$ should be satisfied if $\mathbf w_{\rm c} \neq \mathbf 0$. As such, the achievable rate of decoding $s_k$ at Eve in bps/Hz can be expressed as $R_{\rm p,e\rightarrow \it k} = \log_2\left(1 + \gamma_{\rm p,e\rightarrow \it k}\right) $, where {\small $\gamma_{\rm p,e\rightarrow \it k} = \left|\mathbf v^H\mathbf H_{\rm e}\mathbf w_k\right|^2\Big/\Big( \left|\mathbf v^H\mathbf H_{\rm e}\mathbf w_{\rm c}\right|^2 + \sum_{i\in \mathcal K\backslash\{k\}}\left|\mathbf v^H\mathbf H_{\rm e}\mathbf w_i\right|^2 + {\rm tr}\left(\mathbf H_{\rm e}^H\mathbf v\mathbf v^H\mathbf H_{\rm e}\mathbf Z \right) + \sigma_{\rm e}^2 \Big)$}. Consequently, the achievable SR of LU $k$ in bps/Hz is given by $R_k^{\rm sec} = r_{\rm c,\it k}^{\rm sec} + \left[R_{\rm p,\it k} - R_{\rm p,e\rightarrow \it k} \right]^+$, where $[x]^{+} = \max(x,0)$ and $r_{\rm c,\it k}^{\rm sec}$ denotes the non-negative secrecy common rate allocated to LU $k$. In addition, the non-negative optimization variables $\{r_{\rm c,\it k}^{\rm sec}\}$ need to satisfy the condition $\sum_{k\in\mathcal K}r_{\rm c,\it k}^{\rm sec} \leq R_{\rm c} - R_{\rm c,e}$, which also implies that $R_{\rm c,e} \leq R_{\rm c}$.  

Targeting at maximizing the minimum SR among all the LUs, we formulate the joint design of the transmit precoder and the AN covariance matrix at the AP, the phase shifts at the IRS, and the secrecy common rate allocated to each LU as follows: 
\fontsize{9.7pt}{\baselineskip}
\begin{subequations}\label{RSMA_prob:orig}
	\setlength\abovedisplayskip{4pt}
	\setlength\belowdisplayskip{4pt}
	\begin{eqnarray}
	&&\hspace{-1.3cm}\text{(P1)}: \underset{\mathbf w, \mathbf Z\in\mathbb H^M,\mathbf v, \mathbf r_{\rm c}^{\rm sec}}{\max} \hspace{1mm} \min_{k\in\mathcal K} \hspace{2mm} \left\lbrace r_{\rm c,\it k}^{\rm sec} + \left[R_{\rm p,\it k} - R_{\rm p,e\rightarrow \it k} \right]^+ \right\rbrace \label{RSMA_orig_obj}\\
    \hspace{-3mm}&\text{s.t.}&  \sum\nolimits_{k\in\mathcal K}r_{\rm c,\it k}^{\rm sec} \leq R_{\rm c} - R_{\rm c,e}, \label{RSMA_cons:orig_b}\\
	&& \left\|\mathbf w_{\rm c} \right\|^2 + \sum\nolimits_{k\in \mathcal K}\left\|\mathbf w_k \right\|^2 + {\rm tr}\left(\mathbf Z\right) \leq P_{\text{max}}, \label{RSMA_cons:orig_c}\\
	&& \left|[\mathbf v]_n \right| = 1, \forall n\in\mathcal N, \ [\mathbf v]_{N+1} = 1, \label{RSMA_cons:orig_d}\\
	&& \mathbf Z\succeq \mathbf 0, \ r_{\rm c,\it k}^{\rm sec} \geq 0, \forall k\in\mathcal K, \label{RSMA_cons:orig_e}
	\end{eqnarray}
\end{subequations}
\normalsize where $\mathbf w \triangleq \{\mathbf w_{\rm c}, \mathbf w_1, \cdots,\mathbf w_K\}$ and $\mathbf r_{\rm c}^{\rm sec} \triangleq \{r_{\rm c,1}^{\rm sec}, \cdots, r_{\rm c,\it K}^{\rm sec}\}$. Furthermore, $P_{\max}$ in \eqref{RSMA_cons:orig_c} represents the maximum transmit power at the AP, \eqref{RSMA_cons:orig_d} ensures the unit-modulus constraints on the phase shifts, and \eqref{RSMA_cons:orig_e} imposes the semidefinite and non-negativity constraints on $\mathbf Z$ and $\{r_{\rm c,\it k}^{\rm sec}\}$, respectively. Note that (P1) is an intractable non-smooth and non-convex problem because of the non-smoothness introduced by the operator $[\cdot]^{+}$, the coupled optimization variables in \eqref{RSMA_orig_obj} and \eqref{RSMA_cons:orig_b}, and the non-convex unit-modulus constraints in \eqref{RSMA_cons:orig_d}. Hence, it is challenging, if not impossible, to solve (P1) optimally. 

\vspace{-5mm}
\section{Solution to Problem (P1)}
To facilitate the solution design, we define $\mathbf W_l = \mathbf w_l\mathbf w_l^H$, $\forall l\in\mathcal L \triangleq \{\rm c\}\cup\mathcal K$ and $\mathbf V = \mathbf v\mathbf v^H$, where $\mathbf W_l \succeq \mathbf 0$, ${\rm rank}\left(\mathbf W_l\right)\leq 1$, $\forall l\in\mathcal L$, $\mathbf V\succeq \mathbf 0$, and ${\rm rank}\left(\mathbf V\right)\leq 1$. Besides, we introduce auxiliary variables $t$ and $\mathbf r_{\rm p}^{\rm sec} \triangleq \{r_{\rm p,1}^{\rm sec}, \cdots, r_{\rm p,\it K}^{\rm sec}\}$ and then equivalently convert (P1) into the following form:
\fontsize{9.5pt}{\baselineskip}
\begin{subequations}\label{RSMA_prob:SDR}
	\setlength\abovedisplayskip{4pt}
	\setlength\belowdisplayskip{4pt}
	\begin{eqnarray}
	&&\hspace{-1cm}\text{(P2)}: \hspace{2mm}\underset{\mathcal A}{\max} \hspace{2mm} t\\
	&\hspace{-6mm}\text{s.t.}& \hspace{-2.6mm} r_{\rm c,\it k}^{\rm sec} + r_{\rm p,\it k}^{\rm sec} \geq t, \forall k\in\mathcal K, \label{RSMA_cons:SDR_b}\\
	&& \hspace{-2.6mm} (f_{\rm p,\it k} - g_{\rm p,\it k}) - (f_{\rm e} - g_{\rm p,e\rightarrow \it k}) \geq r_{\rm p,\it k}^{\rm sec}, \forall k\in\mathcal K,\label{RSMA_cons:SDR_c}\\
	&& \hspace{-2.6mm} \sum\nolimits_{k\in\mathcal K}r_{\rm c,\it k}^{\rm sec} \leq (f_{\rm c,\it k} - g_{\rm c,\it k}) - (f_{\rm e} - g_{\rm c,e}), \forall k\in\mathcal K,\label{RSMA_cons:SDR_d}\\
	&& \hspace{-2.6mm} \sum\nolimits_{l\in \mathcal L}{\rm tr}\left(\mathbf W_l\right) + {\rm tr}\left(\mathbf Z\right) \leq P_{\text{max}}, \label{RSMA_cons:SDR_e}\\
	&& \hspace{-2.6mm} [\mathbf V]_{n,n} = 1, \forall n\in\mathcal N\cup\{N+1\},  \label{RSMA_cons:SDR_f}\\
	&& \hspace{-2.6mm} \mathbf Z\succeq \mathbf 0, \mathbf W_l\succeq \mathbf 0, {\rm rank}\left(\mathbf W_l\right)\leq 1, \forall l\in\mathcal L, \label{RSMA_cons:SDR_g}\\
	&& \hspace{-2.6mm} \mathbf V\succeq \mathbf 0, {\rm rank}\left(\mathbf V\right)\leq 1, \label{RSMA_cons:SDR_h}\\
	&& \hspace{-2.6mm} r_{\rm c,\it k}^{\rm sec} \geq 0, r_{\rm p,\it k}^{\rm sec} \geq 0, \forall k\in\mathcal K, \label{RSMA_cons:SDR_i}
	\end{eqnarray}
\end{subequations}%
\normalsize where {\small $\mathcal A \triangleq \{\{\mathbf W_l\in\mathbb H^M\}, \mathbf Z\in\mathbb H^M,\mathbf V\in\mathbb H^{N+1}, \mathbf r_{\rm c}^{\rm sec},\mathbf r_{\rm p}^{\rm sec},t\}$, $f_{\rm p,\it k} \triangleq \log_2\Big(\sum_{i\in \mathcal K}{\rm tr}\left(\mathbf H_k^H\mathbf V\mathbf H_k\mathbf W_i \right) + {\rm tr}\left(\mathbf H_k^H\mathbf V\mathbf H_k\mathbf Z \right) + \sigma_k^2\Big)$, $g_{\rm p,\it k} \triangleq \log_2\Big(\sum_{i\in \mathcal K\backslash\{k\}}{\rm tr}\left(\mathbf H_k^H\mathbf V\mathbf H_k\mathbf W_i \right) + {\rm tr}\left(\mathbf H_k^H\mathbf V\mathbf H_k\mathbf Z \right) + \sigma_k^2\Big)$, $f_{\rm e} \triangleq \log_2\Big(\sum_{i\in \mathcal L}{\rm tr}\left(\mathbf H_{\rm e}^H\mathbf V\mathbf H_{\rm e}\mathbf W_i \right) + {\rm tr}\left(\mathbf H_{\rm e}^H\mathbf V\mathbf H_{\rm e}\mathbf Z \right) + \sigma_{\rm e}^2\Big)$, $g_{\rm p,e\rightarrow \it k} \triangleq \log_2\Big(\sum_{i\in \mathcal L\backslash\{k\}}{\rm tr}\left(\mathbf H_{\rm e}^H\mathbf V\mathbf H_{\rm e}\mathbf W_i \right) + {\rm tr}\left(\mathbf H_{\rm e}^H\mathbf V\mathbf H_{\rm e}\mathbf Z \right) + \sigma_{\rm e}^2\Big)$, $f_{\rm c,\it k} \triangleq \log_2\Big(\sum_{i\in \mathcal L}{\rm tr}\left(\mathbf H_k^H\mathbf V\mathbf H_k\mathbf W_i \right) + {\rm tr}\left(\mathbf H_k^H\mathbf V\mathbf H_k\mathbf Z \right) + \sigma_k^2\Big)$, $g_{\rm c,\it k} \triangleq \log_2\Big(\sum_{i\in \mathcal K}{\rm tr}\left(\mathbf H_k^H\mathbf V\mathbf H_k\mathbf W_i \right) + {\rm tr}\left(\mathbf H_k^H\mathbf V\mathbf H_k\mathbf Z \right) + \sigma_k^2\Big)$, and $g_{\rm c,e} \triangleq \log_2\Big(\sum_{i\in \mathcal K}{\rm tr}\left(\mathbf H_{\rm e}^H\mathbf V\mathbf H_{\rm e}\mathbf W_i \right) + {\rm tr}\left(\mathbf H_{\rm e}^H\mathbf V\mathbf H_{\rm e}\mathbf Z \right) + \sigma_{\rm e}^2\Big)$}. Although (P2) is smooth, it is still non-convex w.r.t. $\mathcal A$. To tackle (P2), we resort to the widely used 
AO method \cite{2019_Miao_Secure,2020_Xinrong_AN,2020_Xianghao_secure_rank1,2021_Hehao_security,2022_Hehao_IRS_security_SWIPT,2021_Zheng_secure_NOMA}. Specifically, we alternatingly optimize $\mathcal A\backslash\{\mathbf V\}$ and $\mathbf V$ until convergence is achieved, with details given below. 

\vspace{-4mm}
\subsection{Optimizing $\mathcal A\backslash\{\mathbf V\}$ for Given $\mathbf V$}
For any given $\mathbf V$, the subproblem w.r.t. $\mathcal A\backslash\{\mathbf V\}$ is given by \looseness=-1
\fontsize{9.5pt}{\baselineskip}
{\setlength\abovedisplayskip{3pt}
\setlength\belowdisplayskip{3pt} 
\begin{equation}\label{RSMA_subprob1}
\underset{\mathcal A\backslash\{\mathbf V\}}{\max} \hspace{1mm} t \hspace{4mm}
\text{s.t.} \hspace{2mm}\eqref{RSMA_cons:SDR_b} - \eqref{RSMA_cons:SDR_e}, \eqref{RSMA_cons:SDR_g}, \eqref{RSMA_cons:SDR_i}.
\end{equation}}%
\normalsize Note that the functions $f_{\rm p,\it k}$, $g_{\rm p,\it k}$, $f_{\rm e}$, $g_{\rm p,e\rightarrow\it k}$, $f_{\rm c,\it k}$, $g_{\rm c,\it k}$, and $g_{\rm c,e}$ are all jointly concave w.r.t. their corresponding variables. Although the concavity of $g_{\rm p,\it k}$, $f_{\rm e}$, and $g_{\rm c,\it k}$ makes the constraints in \eqref{RSMA_cons:SDR_c} and \eqref{RSMA_cons:SDR_d} non-convex, it facilitates the application of the iterative successive convex approximation (SCA) technique \cite{2021_Hao_IRS_RSMA_multiuser}. For ease of notation, let $\mathbf W$, $\hat {\mathbf W}$, and $\tilde {\mathbf W}$ denote the collections of the variables $\{\mathbf W_i\}_{\forall i\in\mathcal L}$, $\{\mathbf W_i\}_{\forall i\in\mathcal K}$, and $\{\mathbf W_i\}_{\forall i\in\mathcal K\backslash \{k\}}$, respectively. Then, given the local feasible points $\tilde {\mathbf W}^r \triangleq \{\mathbf W_i^r\}_{\forall i\in\mathcal K\backslash \{k\}}$ and $\mathbf Z^r$ in the $r$-th iteration, the differentiable concave function $g_{\rm p,\it k}$ is globally upper-bounded by its first-order Taylor expansion, i.e.,
{\fontsize{9.5pt}{\baselineskip}
\setlength\abovedisplayskip{4pt}
\setlength\belowdisplayskip{4pt}
\begin{align}\label{subprob1_sca}
& g_{\rm p,\it k}(\tilde{\mathbf W}, \mathbf Z) \leq g_{\rm p,\it k}(\tilde{\mathbf W}^r, \mathbf Z^r) \nonumber\\
& + \sum\nolimits_{i\in \mathcal K\backslash\{k\}}{\rm tr}\left( \left( \nabla_{\mathbf W_i}g_{\rm p,\it k}\left(\tilde{\mathbf W}^r, \mathbf Z^r\right) \right) ^H\left( \mathbf W_i - \mathbf W_i^r\right) \right) \nonumber\\
& + {\rm tr}\left( \left( \nabla_{\mathbf Z}g_{\rm p,\it k}\left(\tilde{\mathbf W}^r, \mathbf Z^r\right)\right)^H\left(\mathbf Z - \mathbf Z^r\right)\right) \triangleq g^r_{\rm p,\it k}(\tilde{\mathbf W}, \mathbf Z). 
\end{align}}%
\normalsize Similarly, we can obtain the global upper bounds, denoted by $f^r_{\rm e}(\mathbf W, \mathbf Z)$ and $g^r_{\rm c,\it k}(\hat{\mathbf W}, \mathbf Z)$, of $f_{\rm e}$ and $g_{\rm c,\it k}$, respectively, whose expressions are similar to that given in \eqref{subprob1_sca} and are therefore omitted for brevity. Then, by replacing $g_{\rm p,\it k}$, $f_{\rm e}$, and $g_{\rm c,\it k}$ with their respective global upper bounds, we can replace \eqref{RSMA_cons:SDR_c} and \eqref{RSMA_cons:SDR_d} with the following convex constraints: 
{\setlength\abovedisplayskip{4pt}
\setlength\belowdisplayskip{4pt} 
\fontsize{9.7pt}{\baselineskip}
\begin{align}
&(f_{\rm p,\it k} - g^r_{\rm p,\it k}(\tilde{\mathbf W}, \mathbf Z)) - (f^r_{\rm e}(\mathbf W, \mathbf Z) - g_{\rm p,e\rightarrow \it k}) \geq r_{\rm p,\it k}^{\rm sec}, \nonumber\\
& \hspace{6.5cm}\forall k\in\mathcal K, \label{subprob1_sca_cons1}\\
& \sum\nolimits_{k\in\mathcal K}r_{\rm c,\it k}^{\rm sec} \leq (f_{\rm c,\it k} - g^r_{\rm c,\it k}(\hat{\mathbf W}, \mathbf Z)) - (f^r_{\rm e}(\mathbf W, \mathbf Z) - g_{\rm c,e}), \nonumber\\
& \hspace{6.5cm}\forall k\in\mathcal K, \label{subprob1_sca_cons2}
\end{align}}%
\normalsize respectively. However, the rank constraints in \eqref{RSMA_cons:SDR_g} make the problem at hand still non-convex. Thus, we drop the rank constraints by applying semidefinite relaxation (SDR) \cite{2019_Miao_Secure,2020_Xinrong_AN,2020_Xianghao_secure_rank1} that yields 
\fontsize{9.7pt}{\baselineskip}
\begin{subequations}\label{RSMA_subprob1_sca}
	\setlength\abovedisplayskip{4pt}
	\setlength\belowdisplayskip{4pt}
	\begin{eqnarray}
	&& \hspace{-10mm}\underset{\mathcal A\backslash\{\mathbf V\}}{\max} \hspace{2mm} t \\
	&\hspace{-4mm}\text{s.t.}& \eqref{RSMA_cons:SDR_b}, \eqref{subprob1_sca_cons1}, \eqref{subprob1_sca_cons2}, \eqref{RSMA_cons:SDR_e}, \eqref{RSMA_cons:SDR_i}, \mathbf Z\succeq \mathbf 0, \mathbf W_l\succeq \mathbf 0, \forall l\in\mathcal L.
	\end{eqnarray}
\end{subequations}
\normalsize By direct inspection, problem \eqref{RSMA_subprob1_sca} is a convex semidefinite program (SDP) that can be efficiently solved by off-the-shelf solvers such as CVX. Particularly, the adopted SDR is tight for problem \eqref{RSMA_subprob1_sca}. The detailed proof is similar to that of \cite[Theorem 1]{2020_Xianghao_secure_rank1} and we omit it due to the space limitation.  

\vspace{-3mm}
\subsection{Optimizing $\mathbf V$ for Given $\mathcal A\backslash\{\mathbf V\}$}
For any given $\mathcal A\backslash\{\mathbf V\}$, (P2) is reduced to a feasibility-check problem. Inspired by \cite{2019_Qingqing_Joint}, we introduce ``residual'' variables $\Delta t$, $\Delta \mathbf r_{\rm c}^{\rm sec} \triangleq \{\Delta r_{\rm c,1}^{\rm sec}, \cdots, \Delta r_{\rm c,\it K}^{\rm sec}\}$, and $\Delta \mathbf r_{\rm p}^{\rm sec} \triangleq \{\Delta r_{\rm p,1}^{\rm sec}, \cdots, \Delta r_{\rm p,\it K}^{\rm sec}\}$, and then transform the subproblem into the following problem 
\fontsize{9.6pt}{\baselineskip}
\begin{subequations}\label{RSMA_subprob2_trans}
	\setlength\abovedisplayskip{3pt}
	\setlength\belowdisplayskip{4pt}
	\begin{eqnarray}
	&&\hspace{-7mm}\underset{\mathbf V\in\mathbb H^{N+1},\Delta t,\Delta \boldsymbol r_{\rm c}^{\rm sec}, \Delta \boldsymbol r_{\rm p}^{\rm sec}}{\max} \hspace{2mm} t + \Delta t \label{RSMA_subprob2_trans_obj}\\
	&\hspace{-3mm}\text{s.t.}& r_{\rm c,\it k}^{\rm sec} + \Delta r_{\rm c,\it k}^{\rm sec} + r_{\rm p,\it k}^{\rm sec} + \Delta r_{\rm p,\it k}^{\rm sec} \geq t + \Delta t, \forall k\in\mathcal K, \label{RSMA_cons_trans:subprob2_b}\\
	&& (f_{\rm p,\it k} - g_{\rm p,\it k}) - (f_{\rm e} - g_{\rm p,e\rightarrow \it k}) \geq r_{\rm p,\it k}^{\rm sec} + \Delta r_{\rm p,\it k}^{\rm sec}, \nonumber\\
	&& \hspace{5.6cm} \forall k\in\mathcal K, \label{RSMA_cons_trans:subprob2_c}\\
	&& \sum\nolimits_{k\in\mathcal K}r_{\rm c,\it k}^{\rm sec} + \Delta r_{\rm c,\it k}^{\rm sec} \leq (f_{\rm c,\it k} - g_{\rm c,\it k}) - (f_{\rm e} - g_{\rm c,e}), \nonumber\\
	&& \hspace{5.6cm}\forall k\in\mathcal K, \label{RSMA_cons_trans:subprob2_d}\\
	&& \Delta t \geq 0, \Delta r_{\rm c,\it k}^{\rm sec} \geq 0, \Delta r_{\rm p,\it k}^{\rm sec} \geq 0, \forall k\in\mathcal K, \label{RSMA_cons_trans:subprob2_e}\\
	&& \eqref{RSMA_cons:SDR_f}, \eqref{RSMA_cons:SDR_h},
	\end{eqnarray}
\end{subequations}%
\normalsize which is more efficient than the original subproblem when it comes to the converged solution (please refer to \cite{2019_Qingqing_Joint} for a detailed explanation). Similar to problem \eqref{RSMA_subprob1}, the non-convexity of problem \eqref{RSMA_subprob2_trans} stems from the concavity of $g_{\rm p,\it k}$, $f_{\rm e}$, and $g_{\rm c,\it k}$ as well as the non-convex rank constraint in \eqref{RSMA_cons:SDR_h}. As in the previous subsection, we employ the SCA method to tackle this problem. To be specific, by applying the first-order Taylor expansion at the given local feasible point $\mathbf V^r$ in the $r$-th iteration to $g_{\rm p,\it k}$, we obtain $g_{\rm p,\it k}(\mathbf V)  \leq g_{\rm p,\it k}(\mathbf V^r) + {\rm tr}\left( \left( \nabla_{\mathbf V}g_{\rm p,\it k}(\mathbf V^r) \right) ^H\left( \mathbf V - \mathbf V^r\right) \right) \triangleq g^r_{\rm p,\it k}(\mathbf V)$.
Similarly, $f_{\rm e}$ and $g_{\rm c,\it k}$ are globally upper-bounded by their respective first-order Taylor expansions at $\mathbf V^r$, denoted by $f^r_{\rm e}(\mathbf V)$ and $g^r_{\rm c,\it k}(\mathbf V)$, respectively. Accordingly, the non-convex constraints \eqref{RSMA_cons_trans:subprob2_c} and \eqref{RSMA_cons_trans:subprob2_d} can be approximated as
\fontsize{9.6pt}{\baselineskip}
{
\begin{align}
&(f_{\rm p,\it k} - g^r_{\rm p,\it k}(\mathbf V)) - (f^r_{\rm e}(\mathbf V) - g_{\rm p,e\rightarrow \it k}) \geq r_{\rm p,\it k}^{\rm sec} + \Delta r_{\rm p,\it k}^{\rm sec}, \nonumber\\
& \hspace{6.5cm}\forall k\in\mathcal K, \label{subprob2_sca_cons1}\\
& \sum\nolimits_{k\in\mathcal K}r_{\rm c,\it k}^{\rm sec} + \Delta r_{\rm c,\it k}^{\rm sec} \leq (f_{\rm c,\it k} - g^r_{\rm c,\it k}(\mathbf V)) - (f^r_{\rm e}(\mathbf V) - g_{\rm c,e}), \nonumber\\
& \hspace{6.5cm}\forall k\in\mathcal K. \label{subprob2_sca_cons2}
\end{align}}%
\normalsize The only remaining obstacle to solving problem \eqref{RSMA_subprob2_trans} is the non-convex rank constraint in \eqref{RSMA_cons:SDR_h}. Recall that when optimizing $\{\mathbf W_l\}$ in the previous subsection, we dropped the rank constraints and showed the tightness of SDR. However, dropping the rank constraint w.r.t. $\mathbf V$ in problem \eqref{RSMA_subprob2_trans} cannot guarantee a rank-one optimal solution. Thus, instead of applying SDR, we exploit the penalty-based method \cite{2020_Xianghao_secure_rank1} to handle the rank constraint. Specifically, ${\rm rank}\left(\mathbf V\right)\leq 1$ is equivalent to $\left\|\mathbf V\right\|_* - \left\| \mathbf V\right\|_2 \leq 0$. Then, we incorporate the constraint $\left\|\mathbf V\right\|_* - \left\| \mathbf V\right\|_2 \leq 0$ into the objective function \eqref{RSMA_subprob2_trans_obj} by introducing a positive penalty parameter $\rho$ and we replace the non-convex constraints \eqref{RSMA_cons_trans:subprob2_c} and \eqref{RSMA_cons_trans:subprob2_d} with their convex subsets \eqref{subprob2_sca_cons1} and \eqref{subprob2_sca_cons2}, which yields the following problem 
\fontsize{9.5pt}{\baselineskip}
\begin{subequations}\label{RSMA_subprob2_penal}
	\setlength\abovedisplayskip{3pt}
	\setlength\belowdisplayskip{3pt}
	\begin{eqnarray}
	&&\hspace{-2.2cm}\underset{\mathbf V\in\mathbb H^{N+1},\Delta t,\Delta \boldsymbol r_{\rm c}^{\rm sec}, \Delta \boldsymbol r_{\rm p}^{\rm sec}}{\min} \hspace{1mm} - t - \Delta t + \frac{1}{2\rho}\left(\left\|\mathbf V\right\|_* - \left\| \mathbf V\right\|_2\right)  \label{RSMA_subprob2_penal_obj}\\
	&\hspace{3mm}\text{s.t.}& \eqref{RSMA_cons_trans:subprob2_b}, \eqref{subprob2_sca_cons1}, \eqref{subprob2_sca_cons2}, \eqref{RSMA_cons_trans:subprob2_e}, \eqref{RSMA_cons:SDR_f}, \mathbf V\succeq \mathbf 0.
	\end{eqnarray}
\end{subequations}%
\normalsize According to \cite[Proposition 2]{2020_Xianghao_secure_rank1}, problem \eqref{RSMA_subprob2_penal} admits a rank-one solution when $\rho$ is sufficiently small. Note that the convexity of $\left\| \mathbf V\right\|_2$ makes problem \eqref{RSMA_subprob2_penal} still non-convex, which motivates us to replace $\left\| \mathbf V\right\|_2$ by its first-order Taylor expansion-based lower bound. By doing so,  we can approximate problem \eqref{RSMA_subprob2_penal} as 
\fontsize{9.5pt}{\baselineskip}
\begin{subequations}\label{RSMA_subprob2_penal_sca}
	\setlength\abovedisplayskip{-1pt}
	\setlength\belowdisplayskip{3pt}
	\begin{eqnarray}
	&&\hspace{-2.2cm}\underset{\mathbf V\in\mathbb H^{N+1},\Delta t,\Delta \boldsymbol r_{\rm c}^{\rm sec}, \Delta \boldsymbol r_{\rm p}^{\rm sec}}{\min} \hspace{1mm} - t - \Delta t + \frac{1}{2\rho}\bigg(\left\|\mathbf V\right\|_* - \left\| \mathbf V^r\right\|_2 \nonumber\\
	&& \hspace{0.6cm} - {\rm tr}\left(\bm \lambda^r_{\max}\left(\bm \lambda^r_{\max}\right)^H\left(\mathbf V - \mathbf V^r\right)\right)\bigg) \label{RSMA_subprob2_penal_sca_obj}\\
	&\hspace{8mm}\text{s.t.}& \eqref{RSMA_cons_trans:subprob2_b}, \eqref{subprob2_sca_cons1}, \eqref{subprob2_sca_cons2}, \eqref{RSMA_cons_trans:subprob2_e}, \eqref{RSMA_cons:SDR_f}, \mathbf V\succeq \mathbf 0,
	\end{eqnarray}
\end{subequations}
\normalsize where $\bm \lambda^r_{\max}$ is the eigenvector that corresponds to the largest eigenvalue of $\mathbf V^r$. Problem \eqref{RSMA_subprob2_penal_sca} is a convex SDP and existing solvers (e.g., CVX) can be utilized to solve it optimally. 

\vspace{-3mm}
\subsection{Convergence and Computational Complexity Analysis}
The proposed algorithm alternatingly solves problems \eqref{RSMA_subprob1_sca} and \eqref{RSMA_subprob2_penal_sca} until it converges, where the output of one iteration is the input of the next. Notice that due to the adopted series of approximations, the optimal values of problems \eqref{RSMA_subprob1_sca} and \eqref{RSMA_subprob2_penal_sca} serve as lower bounds for those of problems \eqref{RSMA_subprob1} and \eqref{RSMA_subprob2_trans}, respectively, and these lower bounds are tightened monotonically over iterations. In this way, the proposed algorithm is guaranteed to converge to a suboptimal solution of (P2) (and thus (P1)) \cite{2019_Qingqing_Joint,2020_Xianghao_secure_rank1}. Regarding complexity, obviously, it is dominated by solving SDPs \eqref{RSMA_subprob1_sca} and \eqref{RSMA_subprob2_penal_sca} in each iteration. The computational complexity of solving \eqref{RSMA_subprob1_sca} is {\small $\mathcal O\left(\sqrt{M}\log\frac{1}{\varepsilon}\left(KM^3+K^2M^2+K^3\right)\right)$} and that of solving \eqref{RSMA_subprob2_penal_sca} is {\small $\mathcal O\left(\sqrt{N}\log\frac{1}{\varepsilon}\left(N^4+KN^3+K^2N^2+K^3\right)\right)$}, where $\varepsilon > 0$ represents the required solution accuracy in each iteration \cite{2020_Xianghao_secure_rank1}. Hence, the computational complexity of each iteration of the proposed algorithm is about {\small $\mathcal O\Big(\log\frac{1}{\varepsilon}\big(\sqrt{M}(KM^3+K^2M^2+K^3) + \sqrt{N}(N^4+KN^3+K^2N^2+K^3)\big)\Big)$}. 

\begin{figure}[!t]
	\centering
	\vspace{-6mm}
	\subfigbottomskip = -5pt  
	\subfigcapskip = -3.5pt 
	\subfigure[\hspace{-5mm}]{\label{fig:RSMA_vs_Pmax}
		\includegraphics[width = 0.49\linewidth]{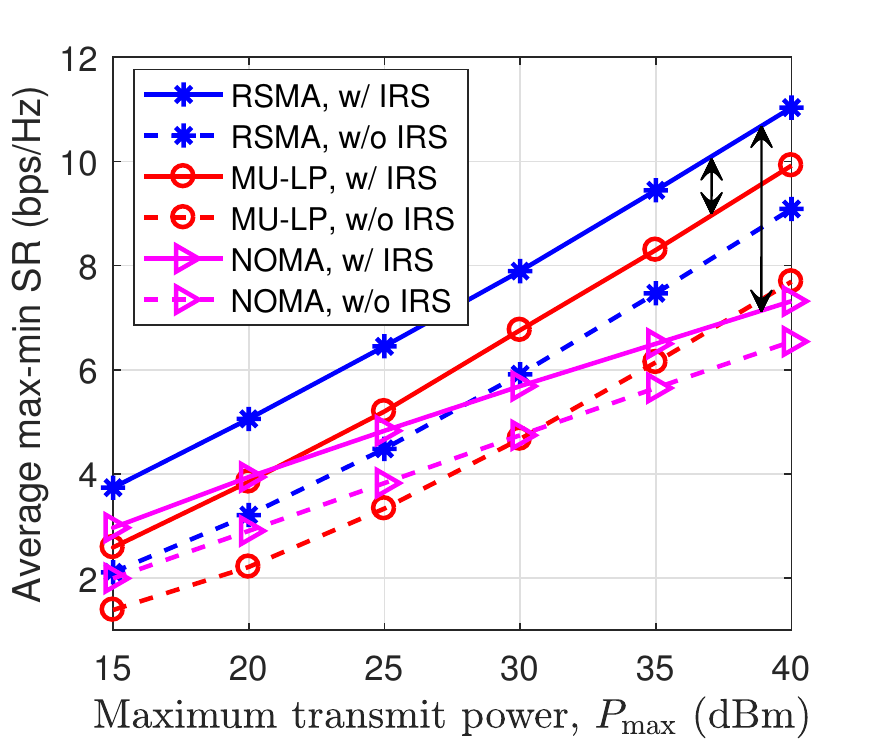}}
	\hspace{-5mm}
	\subfigure[\hspace{-5mm}]{\label{fig:RSMA_vs_P_allocation}
		\includegraphics[width = 0.49\linewidth]{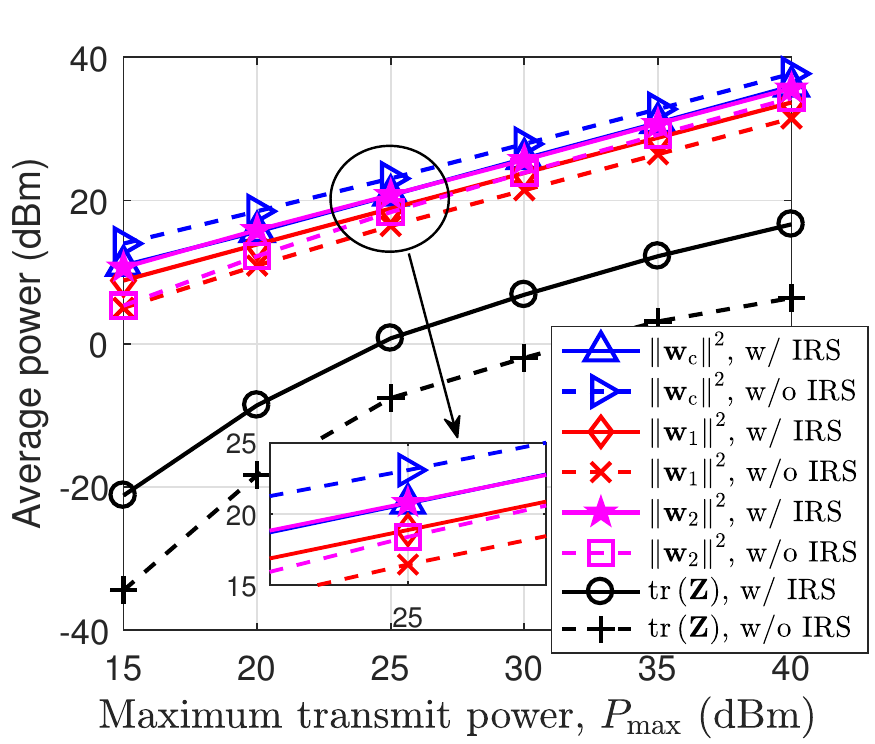}}\\
	\subfigure[\hspace{-5mm}]{\label{fig:RSMA_vs_rate_allocation}
		\includegraphics[width = 0.49\linewidth]{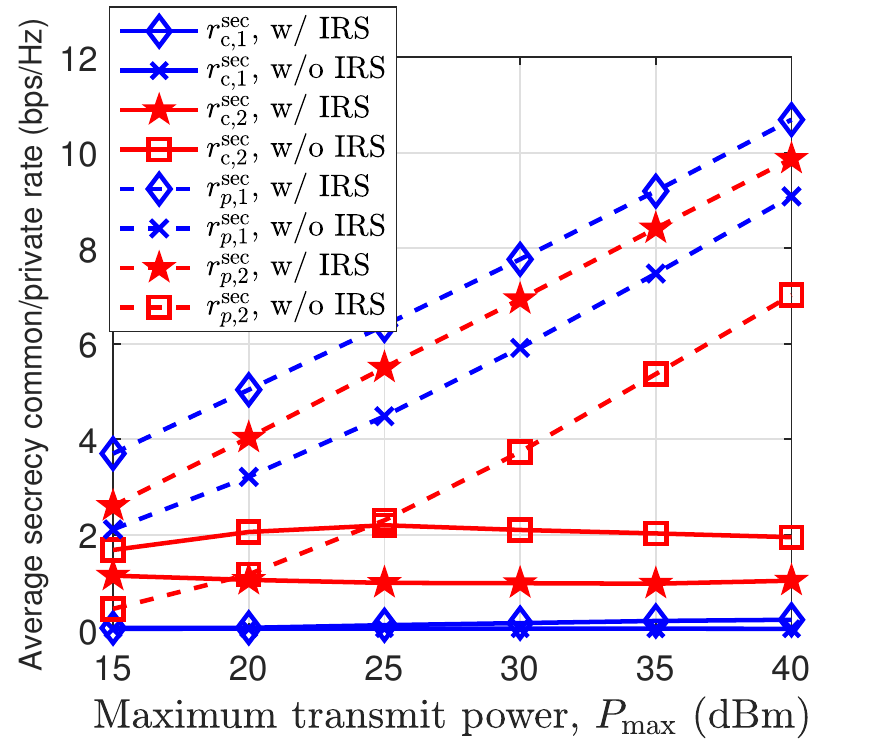}}
	\hspace{-5mm}
	\subfigure[\hspace{-5mm}]{\label{fig:RSMA_vs_N}
		\includegraphics[width = 0.49\linewidth]{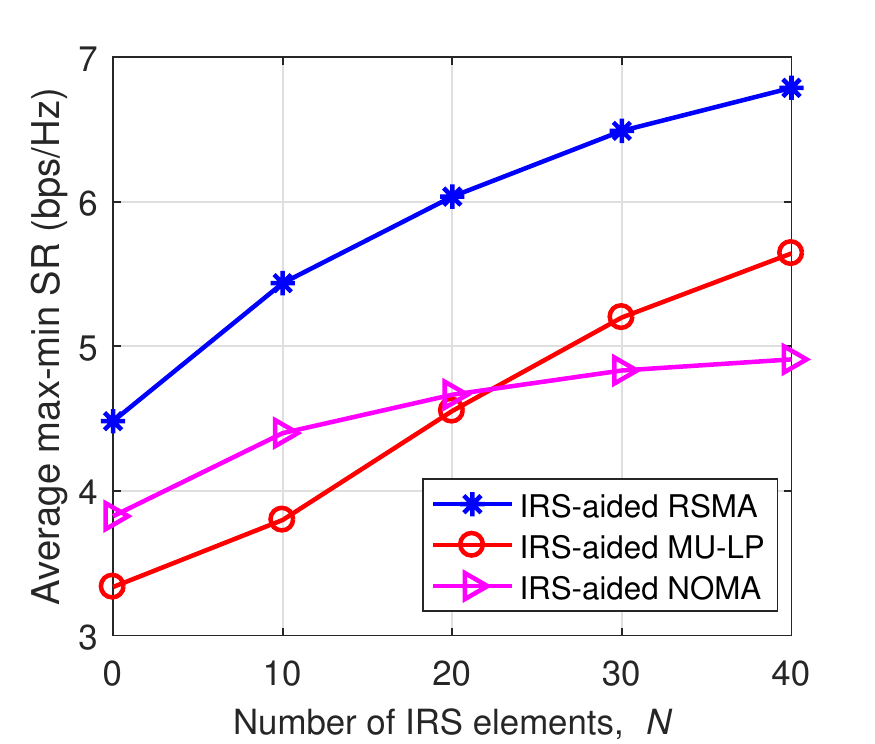}}
	\vspace{1mm}
	\caption{(a) Average max-min SR versus $P_{\max}$; (b) Average transmit power allocation of the RSMA scheme versus $P_{\max}$; (c) Average secrecy common/private rate allocation of the RSMA scheme versus $P_{\max}$; (d) Average max-min SR versus $N$.}
	\label{fig:RSMA_vs_K2}
	\vspace{-6.5mm}
\end{figure}

\vspace{-3mm}
\section{Simulation Results}
This section evaluates the effectiveness of the proposed algorithm via simulations. We consider a two-dimensional (2D) coordinate setup and the locations of the AP, the IRS, and Eve are set as $\left(0,0\right)$, $\left(50,0\right)$, and $\left(45, 0\right)$ in meters (m), respectively. The large-scale path loss model in \cite{2019_Qingqing_Joint} is adopted with the path loss at the reference distance of $1$ m being $-30$ dB and the path loss exponents being $2.2$ for the cascaded links while $3.5$ for the direct links \cite{2022_Hehao_IRS_security_SWIPT}. Furthermore, we adopt the Rician fading model for the cascaded links with a Rician factor of $3$ dB while the Rayleigh fading model for the direct links \cite{2019_Qingqing_Joint}. Unless further specified, other parameters are set as $\sigma_j^2 = -80$ dBm \cite{2019_Qingqing_Joint}, $\forall j\in\mathcal K\cup\{\rm e\}$, $\rho = 5\times 10^{-4}$ \cite{2020_Xianghao_secure_rank1}, and $\varepsilon = 10^{-4}$ \cite{2019_Qingqing_Joint}. All numerical results are obtained by averaging $500$ random channel realizations. For comparison, we adopt two other transmission schemes, namely, MU-LP \cite{2020_Xianghao_secure_rank1,2021_Hehao_security,2022_Hehao_IRS_security_SWIPT} and NOMA \cite{2021_Zheng_secure_NOMA}. \looseness=-1  

Fig. \ref{fig:RSMA_vs_Pmax} depicts the average max-min SR of different strategies versus $P_{\max}$. Here, we set $M = 2$, $N = 30$, and $K = 2$ with LUs $1$ and $2$ being located at $\left(0, 20\right)$ and $\left(50, 5\right)$ in m, respectively. Two cases with and without the IRS are considered. As can be seen, the IRS-aided designs are obviously superior to their counterparts without the IRS. This is expected since the IRS phase shifts can be properly adjusted not only to enhance the signal strength at the LUs while degrading that at Eve, but also to decrease the interference power at the LUs while increasing that at Eve. It is also observed that our proposed IRS-aided RSMA scheme significantly outperforms the other two strategies. To acquire more insights, we plot the transmit power allocation and the secrecy common/private rate allocation obtained by the RSMA scheme in Figs. \ref{fig:RSMA_vs_P_allocation} and \ref{fig:RSMA_vs_rate_allocation}, respectively. Fig. \ref{fig:RSMA_vs_P_allocation} shows that the transmit power allocation changes greatly after the introduction of the IRS, which corroborates that the IRS is capable of reconfiguring the wireless propagation environment. 
From Fig. \ref{fig:RSMA_vs_rate_allocation}, it can be seen that since the obtained secrecy private rate of LU $2$ is much smaller than that of LU $1$, almost all of the achievable secrecy common rate is allocated to LU $2$ to improve its total SR, thereby ensuring the secrecy fairness between LUs $1$ and $2$. Moreover, we note that with increasing $P_{\max}$, the secrecy private rates increase rapidly, while the secrecy common rates change slightly. A possible explanation is that as Eve's received power from $\mathbf w_{\rm c}$ becomes significant in the high $P_{\max}$ regime, the common message contributes little to the total achievable secrecy common rate. Yet, it can serve as AN and contribute much to the individual secrecy private rates. 
It is also worth mentioning that by plotting the received power from different signals at the LUs and Eve in the RSMA scheme, we find that the IRS can enhance the signal strength of all the LUs, mitigate the multi-LU interference, and enhance the AN/common message power at Eve at the same time. However, the corresponding figures are unable to be presented here due to the space limitation. 
Besides, we observe from Fig. \ref{fig:RSMA_vs_Pmax} that in the case without IRS, the max-min SR of the MU-LP scheme is noticeably higher than that of its NOMA counterpart when $P_{\max}$ is larger than about $33$ dBm. This is consistent with the result in \cite{2021_Bruno_NOMA_SDMA}, which states that for two-user MISO broadcast channels, MU-LP strictly outperforms NOMA at high signal-to-noise ratio (SNR) as the max-min fair (MMF) multiplexing gain of MU-LP is twice that of NOMA. When deploying an IRS, however, the MU-LP scheme performs better than the NOMA scheme when $P_{\max}$ is larger than about $23$ dBm. The reason is that the IRS can lead to a high SNR even when $P_{\max}$ is not that large. 

\begin{figure}[!t]
	\vspace{-4mm}
	\setlength{\abovecaptionskip}{-3pt}
	\setlength{\belowcaptionskip}{-1pt}
	\centering
	\includegraphics[width=0.23\textwidth]{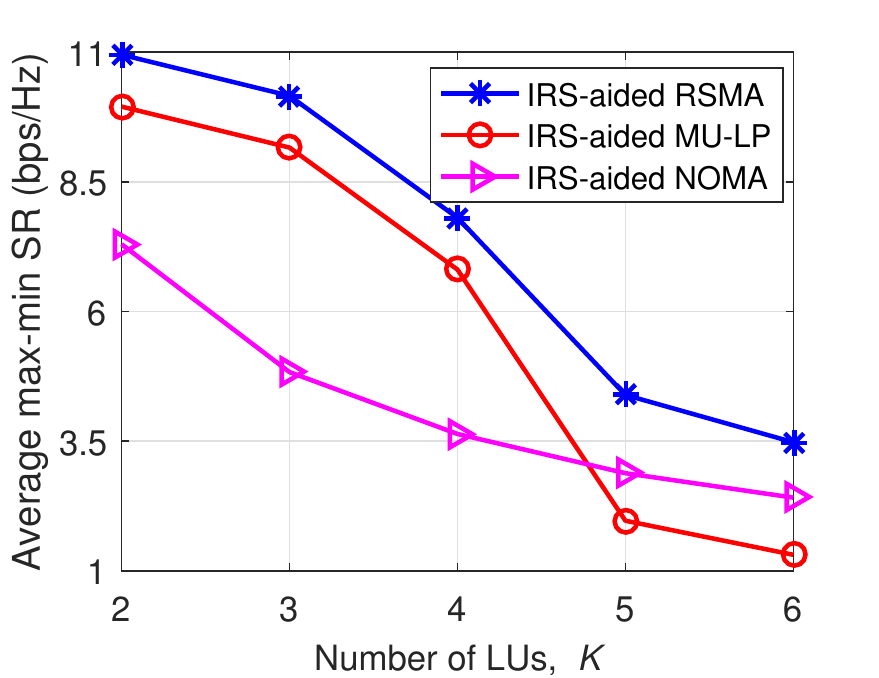}
	\caption{Average max-min SR versus $K$.}
	\label{fig:RSMA_vs_K}
	\vspace{-6mm}
\end{figure}

Following the same setup in Fig. \ref{fig:RSMA_vs_Pmax}, we investigate in Fig. \ref{fig:RSMA_vs_N} the impact of the number of IRS elements on the system SR performance with $P_{\max} = 25$ dBm. It is seen that the proposed IRS-aided RSMA scheme always obtains the highest max-min SR. This is understandable since $1$-layer RS is always a super-scheme of MU-LP regardless of the value of $K$ while NOMA is a sub-scheme of $1$-layer RS in a two-user case (although it is not the case when $K>2$) \cite{2021_Bruno_NOMA_SDMA}. It is also notable that the performance of the IRS-aided MU-LP scheme exceeds that of its NOMA counterpart when $N$ becomes large. This is because increasing $N$ with a fixed $P_{\max}$ can be viewed as increasing $P_{\max}$ with a fixed $N$ that results in an increase in SNR. Finally, we observe that when a certain level of performance is desired, the surface size of the IRS required by the RSMA scheme is much smaller than that required by the other two schemes. This suggests that the IRS-aided RSMA scheme is more appealing to space-limited scenarios. \looseness=-1

Fig. \ref{fig:RSMA_vs_K} plots the average max-min SR of different schemes versus $K$ with $M = 4$, $N = 20$, and $P_{\max} = 35$ dBm. The locations of LUs $1$ and $2$ are identical to those set in Fig. \ref{fig:RSMA_vs_Pmax} and LUs $3 - 6$ are located at $\left(0,-20\right)$, $\left(50,-5\right)$, $\left(55,0\right)$, and $\left(-20,0\right)$ in m, respectively. It is observed that the proposed IRS-aided RSMA scheme still performs the best in both considered underloaded and overloaded regimes. This observation is in line with the theoretical and numerical results in \cite{2021_Bruno_NOMA_SDMA} that $1$-layer RS always achieves the same or higher MMF multiplexing gains and rates than MU-LP and NOMA. Another observation is that when $K$ increases from $4$ to $5$, the max-min SR achieved by the IRS-aided MU-LP scheme drops sharply and is no longer higher than that achieved by the IRS-aided NOMA scheme. This is because the MMF multiplexing gain of MU-LP is $1$ if $K\leq M$ and $0$ otherwise, while that of NOMA is $1/K$ whenever $K \geq 2$ \cite{2021_Bruno_NOMA_SDMA}.  

\vspace{-1.8mm}
\section{Conclusion}
\vspace{-0.8mm}
This letter investigated the potential performance gain of integrating RSMA with IRS in a secure communication system. A computationally efficient iterative algorithm was developed to guarantee the secrecy fairness among the LUs, where the transmit/reflect beamforming with AN and the secrecy common rate allocation were jointly optimized. Simulation results showed that our proposed IRS-aided RSMA strategy can achieve a higher secrecy performance enhancement than the existing IRS-aided MU-LP and NOMA schemes. Possible extensions include the case of imperfect CSI \cite{2022_Hehao_IRS_security_SWIPT}, adopting the generalized RS strategy \cite{2018_YiJie_RSMA}, and investigating whether AN is needed in a secure RSMA-based system (especially in the case of multiple Eves), which are left for future work. \looseness=-1 

\vspace{-2mm}
\bibliographystyle{IEEEtran}
\bibliography{rate-splitting}

\end{document}